\documentclass[onecolumn, prl, showkeys, nofootinbib]{revtex4}

\usepackage{ifthen}
\newboolean{anon}
\setboolean{anon}{false} 

\usepackage[%
	pdfstartview = {FitH},%
    pdftitle = {{Plagiarism: Words and ideas}},%
 	pdfauthor = {{Mathieu Bouville}},%
	pdfsubject = {{Plagiarism is a crime against academy. It deceives readers, hurts plagiarized authors, and gets the plagiarist undeserved benefits. However, when one asks precise questions, the cut and dry situation becomes less obvious.}},%
	pdfkeywords = {{academic dishonesty; academic integrity; academic misconduct; cheating; copyright infringement; ethics; intellectual property; research misconduct}},%
    hyperindex,%
    bookmarksopen,%
    bookmarksopenlevel=1,%
    bookmarksnumbered,%
    breaklinks
]{hyperref}

\begin{document}

\addcontentsline{toc}{Chapter}{Plagiarism: Words and ideas}
\title{Plagiarism: Words and ideas}

\ifthenelse{\boolean{anon}}{}{
\author{Mathieu Bouville}
	\email{m-bouville@imre.a-star.edu.sg}
	\affiliation{Institute of Materials Research and Engineering, Singapore 117602}
	\affiliation{Institute of High Performance Computing, Singapore 117528}
}

\begin{abstract}
Plagiarism is a crime against academy. It deceives readers, hurts plagiarized authors, and gets the plagiarist undeserved benefits. However, even though these arguments do show that copying other people's intellectual contribution is wrong, they do not apply to the copying of words. Copying a few sentences that contain no original idea (e.g.\ in the introduction) is of marginal importance compared to stealing the ideas of others. The two must be clearly distinguished, and the `plagiarism' label should not be used for deeds which are very different in nature and importance.
\end{abstract}
\keywords{academic dishonesty; academic integrity; academic misconduct; cheating; copyright infringement; ethics; intellectual property; research misconduct}
\maketitle

\vspace{0mm}
{\bf\noindent Article published by \emph{Science and Engineering Ethics} ---
\textsc{doi}:\ 10.1007/s11948-008-9057-6}
\vspace{0mm}

\section{Introduction}
Plagiarism is ``the worst of bad behaviour''~\cite{Maddox-95}, ``a sin''~\cite{Loui-02}, ``academic high treason''~\cite{Hexham-92}, etc. In fact, \citet{Park-03} amused himself gathering a lot of such ``rhetoric of plagiarism,'' which he finds ``nothing if not colourful.'' Naturally, calling plagiarism a hateful blasphemy is not the same as showing that it is wrong.

\subsection{The unbearable vagueness of plagiarism}
The word `plagiarism' is applied to vastly different situations, for crimes as well as misdemeanors and even for deeds of unclear wrongness: ``from sloppy documentation and proof-reading to outright, premeditated fraud. Few other terms that we commonly use in our classes have such widely differing meanings''~\cite{Wilhoit-94}. One may call all instances `plagiarism' and distinguish between minor and major ones~--- as one calls both killing in self-defense and genocide `killing' without treating them as equally blameworthy. Alternatively, one may choose to use the word `plagiarism' only for cases where the author obtained unwarranted benefits ---maybe calling other instances ``poor academic practice''~\cite{Burkill-04}---, just like one uses the word `murder' for certain kinds of manslaughter only; for example, to \citet{Yilmaz-07} ``borrowing sentences in the part of a paper that simply helps to better introduce the problem should not be seen as plagiarism.'' In any case, it is important to be consistent. Unfortunately few authors are thus cautious; for instance \citet{Maddox-95} aims to show that ``plagiarism is worse than mere theft'' (his title) but focuses on articles that merely used ``background material'' from others.

\Citet{Giles-06} writes that ``the search [of Sorokina] turned up 677 examples of possible plagiarism.'' More precisely, \citet{Sorokina-06} found ``677 pairs of documents with at least four sentences sharing uncommon 7-grams'' (i.e.\ sentences with at least seven consecutive words in common). One should note that 80\% of these have six or fewer sentences in common: then only a paragraph was taken from another's work. Out of the twenty pairs of articles \citet{Sorokina-06} inspected by hand four were false positives. They also mention that ``many of the isolated copied sentences are of `background' nature, containing neither particularly unique information content nor stylistic virtue.'' Moreover, 677 is not a number of articles but of pairs of similar articles: an article using multiple sources would be counted as many times. In other words, this number does not mean anything. (On top of using a number that is plainly misleading, \citet{Giles-06} mentions Sorokina's name without providing a reference to her article~\cite{Sorokina-06}.)

\subsection{Plagiarism: words or ideas?}
Plagiarism (if explicitly defined at all) is generally taken to mean the appropriation of the words and ideas of others. However, the specific status of words and ideas is not always made clear. After talking about ``the theft of other people's creativity and of their conception of what the world is like'' and about winning ``credit for notions that are not original,'' \citet{Maddox-95} writes that plagiarism ``can be dealt with by the simple question, `Have the words been copied, or have they not?'\,''~--- this seems to indicate that ideas are what plagiarism is about and words are how plagiarism is detected. However, he then assumes that copying words that include no original idea (e.g.\ ``background material'') is plagiarism.

Even though in canonical cases of plagiarism the dichotomy between words and ideas is moot (if I copy a whole article I copy both words and ideas, so that which of the two `thefts' creates the offense is unclear), it is possible to copy a part of an article that contains no original idea (e.g.\ the background section) or to take ideas but not words, by paraphrasing. The status of words and ideas must be clarified to precisely define plagiarism, which is important for both conceptual and practical reasons. 
An important difference between copying words and copying ideas is the kind of argument that applies to them (a given argument is unlikely to show both to be wrong). This is generally overlooked and arguments are proposed against plagiarism that are in fact relevant only to the copying of original ideas.

\subsection{Pseudo-plagiarism}
Certain practices are labeled plagiarism, even though they are different concepts, e.g.\ 
``using `blanket' references, `second-generation' references, and duplicate or repetitive publication of one's own previously published work''~\cite{Skandalakis-04}. That these are not plagiarism does not mean that they are not wrong, but simply that they require separate arguments against them.

\emph{Vague definition.} In some cases, the issue is the vagueness of the definition of `plagiarism.' For example, dealing with a quotation with a citation (but no page number) but without quotation marks, \citet{Hexham-92} claims: ``this is an example of plagiarism [\,\ldots]\ because appropriate quotation marks are not used nor are we given a page reference to the source''~--- even though the absence of a page number has nothing to do with plagiarism.

\emph{Self-plagiarism.} The publication of the same (or a very similar) article in multiple journals is often labeled `self-plagiarism' and held as very wrong. (\Citet{Benos-05} even dedicate more space to this practice than to plagiarism,  falsification, and fabrication combined.) 
\Citet{Giles-05} writes that plagiarism means ``attempting to pass off someone else's work as your own. Duplicate publication, or self-plagiarism, occurs when an author reuses substantial parts of their own published work without providing the appropriate references.'' Obviously, this is not plagiarism based on his own definition~--- and a definition that is in the previous sentence! 
(Even though \citet{Hexham-92} is right that ``to avoid confusion here perhaps it is better to drop the term `self-plagiarism','' one should understand that the reason why people call this `self-plagiarism' is precisely to create confusion: no argument is necessary when something has been labeled `plagiarism,' the label suffices.)

\emph{``Plagiarism of secondary sources.''} \Citet{Martin-94} claims that ``a more subtle plagiarism occurs when a person gives references to original sources, and perhaps quotes them, but never looks them up, having obtained both from a secondary source~--- which is not cited (\citet[456--457]{Bensman-88}). This can be called plagiarism of secondary sources.'' If I read in~[101] a quotation from~[100] and re-use the quotation (with quotation marks and citing [100]), I do not plagiarize~[100] because I cite it and I do not plagiarize~[101] because I do not use its words or its ideas. If no work has been plagiarized then there is obviously no plagiarism.%
\footnote{All arguments about citing a work one did not read seem to assume that the work was not read out of laziness or some other character flaw of the citing author. But if I learn about an article that is highly relevant to my work but cannot read it (e.g.\ my library does not have it and cannot get it) I can either not cite it (but one can then accuse me of not citing a work that I knew to be relevant) or cite it without reading it. Neither is satisfactory of course, but there may not be a third choice. Such deadlock situations are common and it is also common to blame people for not making the right decision, even when no outcome seems acceptable~\cite{author-dilemma, author-cheating}. 
One should also note that it is possible to know a work without reading the book~--- that I never read Newton's \emph{Principia Mathematica} does not mean that I do not know Newtonian mechanics.}

\subsection{Notes on scope}
In the remainder of this article, I will deal only with real plagiarism, i.e.\ plagiarism of others.
I will focus on academic work, such as books and journal articles (I will not consider the issue of plagiarism in the contexts of fiction, music, etc.). 
Unless otherwise noted, I will be concerned with the plagiarism of researchers rather than with that of students. 
I will look at plagiarism from an ethical rather than legal viewpoint, i.e.\ I will ask what is wrong rather than what is illegal (unlike \citet{Standler-00} for instance).

\section{Copying words without copying ideas}

\subsection{The relative importance of words and ideas}
\emph{Facts and theories, not words, are the core of science.}
Scholarship means creating new knowledge. Words are only a means to communicate this knowledge, the knowledge itself is made of facts, concepts, etc. 
A scientist whose ideas are weak is certainly not a good scientist; a scientist whose English is weak can be a good scientist.%
\footnote{If the writing is especially beautiful, witty, entertaining then by using someone else's words I would steal some of the art or of the flair of the original author. But in many disciplines, beauty, wit, and entertainment are rare and not valued (if not banned) so that this counter-argument is limited to fields that value writing and to authors who actually write well.}
An experimental result that is described using different words is not a different result and its scientific importance is not affected by the wording. 
This can be contrasted with poetry for instance: words are what poetry is all about, ``the wording is the essence of the novelty''~\cite{Vessal-07}. A poet who does not have a way with words is a poor poet. A poem that is paraphrased is a different poem, the two may have vastly different merits.

Since the core of science are facts and theories, not words, new experimental results and new models are the basis of new scientific knowledge. \Citet{Yilmaz-07} is right when he writes: ``our results are [original]~--- and these are the most important part of any scientific paper.'' As notes \citet{McCutchen-94}, ``the thing of value that is stolen is intellectual product.''

\emph{Sacrificing words rather than ideas.}
\Citet{Williams-07} mentions authors who do not ``have English as their mother tongue, and struggle to represent the background to their work in good English [\,\ldots]\ 
many such authors succumb to the temptation to use eloquent phrases, sentences or even whole paragraphs found in recently published papers in order to improve their own work.''%
\footnote{One can remark that this offense can be committed only by non-native speakers (\emph{qua} non-native speakers). Arguments against plagiarism are based on desert (the plagiarist does not deserve, the plagiarized would have deserved), but do native speakers deserve the advantage hereby enforced? Also, it is easy to set constraints that will not apply to oneself and to consider that problems other people may face cannot be serious.} 
 According to one such author this is not plagiarism, ``just borrowing better English''~\cite{Yilmaz-07}. These authors might in fact be praised, who are ``disinclined to sacrifice quality and accuracy for want of linguistic expertise''~\cite{Vessal-07}. Science would be more hurt by an incomprehensible article than by copied words: if the article is so unclear that the intellectual contribution is lost then arguing against this practice requires to hold words as more important than ideas~--- better lose ideas than copy words. No one would seriously thus~argue.

\emph{On paraphrasing.}
One may argue that, in any case, the author should paraphrase the work of others rather than copy it \emph{verbatim}. But this requires that words be unimportant: if words were really important then paraphrasing would be impossible (this is the case with literature~--- a poem cannot be paraphrased because a paraphrased poem would be a whole new poem, or perhaps no poem at all). Someone saying that authors must paraphrase thereby says that words are unimportant.

\subsection{Negative consequences of plagiarism}
A number of arguments have been used against plagiarism. However, they do not always apply to what authors wish to prove wrong. I will present these arguments and ask whether they are relevant to cases where only words are copied, not an original intellectual contribution. 

\emph{Hurts the plagiarized author.}
A common argument against plagiarism is that plagiarized authors may lose citations and recognition, which in turn may lose them promotions or awards~--- if no one knows that they are the authors of important work, they cannot get credit for it. They may also lose market shares and royalties in the case of a book. In other words, the plagiarized authors do not receive what they would have deserved to receive. 
One should nonetheless note that this applies to the cases of plagiarists using someone else's ideas, not words~--- in most disciplines one does not get recognized for one's words but for one's ideas (there are exceptions, such as literature). Fame and promotion are not based on the quality of the English in one's background sections~--- I have never heard of an endowed chair of background section. Moreover, had the plagiarists paraphrased the paragraph they copied \emph{verbatim} they would not have cited the original paper (and would have had no reason to): the original authors cannot lose something (e.g.\ citations) that they would not have received anyway.

\emph{Undeserved benefits for the plagiarist.}
To scientists, plagiarism is akin to ``the commercial crime of `passing off', by which a manufacturer may seek to claim a distinguished brand-name for inferior goods''~\cite{Maddox-95}. Concretely, ``the thieving author win[s] credit for notions that are not original''~\cite{Maddox-95} and receives awards and promotion for the work of others. This is related to the previous argument in that the plagiarist steals citations, fame, royalties, etc.\ from the plagiarized~--- what one gains the other loses. Once again, this applies only to intellectual contribution, not mere words: a plagiarist will not gain much by copying \emph{verbatim} a paragraph that contains no original~idea.

\emph{Negative consequences for the reader.}
Two negative consequences of plagiarism are that ``readers are unable to reconstruct the route by which [ideas] have come to see the light of day''~\cite{Maddox-95} and that ``when citations are left out of documents, the reader is deprived of one of the most fruitful ways of seeking additional resources related to the paper topic''~\cite{Snapper-99}. Obviously, these arguments apply to the copying of ideas, not of mere words.

\emph{Breaks Trust.}
\Citet{Hinman-02} remarks that ``scholarship and research are strongly dependent on trust,'' which plagiarism may jeopardize. However, some hold that plagiarism ``is not a significant threat to confidence in science''~\cite{Stenflo-04}. In any case, the risk of plagiarism is not the only reason why I may not trust the work of others: for instance articles may contain errors because the authors were careless. If I were to choose between the two, I would rather have the authors never be careless rather than not try to pass off as theirs the work of others: the first issue is just far more common (plagiarism concerns only 0.02\%~\cite{Claxton-05}, 0.04\%~\cite{Errami-08}, or 0.2\%~\cite{Sorokina-06} of all articles). 

Some may not trust authors who copied background material: who knows if they did not also copy something else. One should note that if copying background were widely accepted while other plagiarisms were not then one would not even think of such a possible correlation. After all, we do not say `he said he killed him in self-defense but how can we know that he would not kill someone else in cold blood?'. This is because we believe that killing in self-defense really is different from murder. Copying background material may be harmful (through loss of trust) only if people believe that it is: this is self-fulfilling.

\subsection{Deception}
One may argue that, even in the absence of a clear loss on the side of the readers or of the plagiarized authors, plagiarism is still a ``deliberate attempt to deceive the reader''~\cite{Hexham-92}.%
\footnote{One may note that, while in the case of researchers who do not cite sources one may mistakenly believe that the authors obtained the data themselves, such confusion is unlikely with undergraduate work. Picture an anthropology student writing an essay on some remote people in which citations are missing. Someone claiming that this student tried to deceive the teacher would thereby claim that the teacher may thus be deceived, i.e.\ made to believe that the student gathered the data himself.} 
However, it is noteworthy that readers cannot be deceived \emph{tout court}: deception means that readers are made to believe something that is false, and one must specify what this thing is. In basic plagiarism, they are made to believe that the names listed on the article are those of the actual authors, and in the case of fabrication, that the experiments were actually done. If authors copied a paragraph in their introduction the readers will wrongly believe that every single word in the article was from the authors. But no reader would feel betrayed upon learning that the introduction has been improved by a copy-editor, i.e.\ someone who is not listed as author. If the alteration resulting from the lifting of a few sentences from the work of others is not greater than the alteration from copy-editing, one cannot claim that the readers are deceived by one but not by the other. (Of course copy-editors agree to this state of affairs whereas a plagiarized author does not, but this is not a matter of deception of the readers.) 
 
If one were to ask why copying a few sentences of background material is wrong, the reply would be that it is a form of plagiarism. And if one follows up asking why plagiarism is wrong, one would receive an answer (deception of the readers, unfair advantage of the plagiarist, or disadvantage of the plagiarized) that would not apply in this case. The only thing that may make the copying of a few sentences that contain no original intellectual contribution deeply wrong is the `plagiarism' label: if one tried to show this to be very wrong without referring explicitly to plagiarism, one would fail.
Even though these arguments can show that copying someone else's intellectual contribution is wrong, they are irrelevant to the copying of words that do not contain original intellectual contribution. ``Copying [words] creates a witness to crime, but is not necessarily crime itself''~\cite{McCutchen-94}.

\subsection{Evil or trivial? A criterion}
Plainly, there is a great difference between an article that must be withdrawn because it is copied \emph{verbatim} and an article that needs a two line erratum to clarify that a given sentence was a direct quotation from the source mentioned. In fact, \citet{Clarke-06} worked out an extensive set of criteria to judge the importance of a given instance of plagiarism. He distinguishes between ``extreme forms of plagiarism [such as] `plagiarism of authorship' or `appropriation of entire works'\,'' and the ``many instances of plagiarism [that] are misjudgments or errors that are appropriately addressed in simple ways.'' 
To \citet{Burkill-04}, ``there is a fine line between plagiarism and poor academic practice.'' (They add that ``it is sometimes difficult to be sure that you are on the right side of that line,'' without specifying which of plagiarism and ``poor academic practice'' is right.) 

Some will say that authors must follow the guidelines of the style manuals, and plagiarists obviously do not follow those on proper citation. But such manuals only state was is correct and was is not, they are not (and have no business being) concerned with establishing the relative wrongness of different faults~--- `plagiarism is wrong because it is incompatible with stylistic rules' is on a par with `using too many commas is wrong because it is incompatible with stylistic rules.' For this reason, one cannot rely on style manuals to argue that failure to cite properly is incommensurably worse than bad punctuation or other trifles. 

In any case of plagiarism, one should ask what would have happened had the author cited properly. If the work has been copied \emph{verbatim} from another's book or article, a proper citation would have been along the lines of `the whole thing was written by someone else, I did nothing.' But if the author only uses small parts from a given work and if these do not contain any original idea, the result of proper citation would have been a work of very similar merit. I propose this as a criterion to judge how wrong a certain instance of plagiarism is: if including proper citations would have been a fatal blow this instance is clearly plagiarism and clearly wrong; otherwise, that a work is (akin to) plagiarism does not mean that it should be withdrawn and the author fired. If proper citation would have had little impact on the paper, lack thereof should also have little impact. (If it is possible to rewrite an article ---without changing its substance--- to make some citation unnecessary then only words had be borrowed; otherwise, one must be using an idea from this work and a citation is required.)

\section{Plagiarism hunting}

\subsection{String-based criteria for plagiarism}
Plagiarism has been equated with strings of two--three~\cite{Rathus-93}, four~\cite{Hexham-92}, or seven~\cite{Sorokina-06} consecutive words that are found in two different articles. 
In his definition of plagiarism, \citet{Hexham-92} writes: ``Academic plagiarism occurs when a writer repeatedly uses more than four words from a printed source without the use of quotation marks and a precise reference to the original source.'' No rationale is given for choosing this particular number (or why plagiarism can be defined as strings of words of a certain length in the first place). 
One reason why people pay so much attention to copied words is that this is easy to detect (a computer can do it). One simply asks ``\,`Have the words been copied, or have they not?' That means that when a plagiarist is unmasked, the case appears open and shut. So it is no wonder that plagiarism is often the most easily punished of academic misdemeanours''~\cite{Maddox-95}. Establishing plagiarism of ideas is less straightforward~\cite{Weyland-07}.

Relying on strings of words and on software raises a number of issues.

\emph{Nature of the definition of plagiarism.}
\Citet{Sorokina-06} chose strings of seven words because they found that strings of six words included too many false positives and strings of eight words missed some actual cases of plagiarism.  
What their work thus shows is that identical strings of seven words tend to correlate with plagiarism. 
But a risk is that `identical strings of seven words are a hint that plagiarism may have occurred' may become `plagiarism occurs when there are identical strings of seven words.' In other words, plagiarism would become defined in terms of strings of words (\citet{Hexham-92} for example is guilty of such a drift), even though \citet{Sorokina-06} had texts inspected by humans to decide what qualified as plagiarism, based on an independent definition of plagiarism. Such a procedure simply cannot generate a definition of plagiarism. And in fact it makes no sense to \emph{define} plagiarism in terms of strings of words~--- in particular, showing that plagiarism is wrong would then require to show that copying seven consecutive words is wrong, but copying six consecutive words is not.

\emph{Lack of effect on plagiarism.}
According to \citet{Bechhoefer-07}, ``every paper submitted to arXiv could be examined by a search engine that looks for overlap or correlation with all previous arXiv submissions. [\,\ldots] And although plagiarists might opt to copy and translate from foreign-language journals, or simply alter wording enough to pass muster, making it more difficult will at least discourage the lazier offenders.''
Copying and pasting an article takes fives minutes. Paraphrasing a whole article to avoid detection may take five hours. This is still far shorter than the five months it would take to do the research. The driving force for plagiarism would still be huge. The main result of the systematic use of such software will be to make plagiarists hide better, it will have little effect on the amount of academic plagiarism.

\emph{Presumption of guilt.}
\Citet{Hinman-02} complains: ``The automatic submission of papers to an anti-plagiarism site strikes me as akin to mandatory urine testing for athletes, a sign precisely of lack of trust.'' \Citet{Townley-04} likewise point out that this would ``set up a distrustful relationship.'' They add that ``alongside the requirement to check the proliferation of plagiarism, however, we must also consider the collateral effects of the strategies we employ.'' While treating everyone as potentially guilty is never satisfying it may be a necessary evil in case of an epidemic (e.g.\ doping in professional sport). An important difference between athletes and scholars (but not students) is that world-class athletes have used forbidden products in spite of the controls, whereas plagiarism concerns very few scholars in spite of the absence of systematic control (serious plagiarism is found in one article out of thousands~\cite{Claxton-05, Errami-08, Sorokina-06}, and there is ``an even smaller percentage of authors [concerned], since many [instances of plagiarism] come from repeat offenders''~\cite{Sorokina-06}). Moreover, unlike athletes with doping, ``prominent (highly cited) authors [\,\ldots]\ do not appear to reuse text from others''~\cite{Sorokina-06}. 

\emph{Plagiarism, no plagiarism, and that is all.}
Another issue is that plagiarism may become even more a binary issue than it already is~--- there are two cases, plagiarism and no plagiarism, with no degree. \Citet{Sorokina-06} describe precisely the method they use, mentioning that they are looking for ``($\mu$, $L$; $t$, $m$)-similar documents'' using values for certain parameters that are rather arbitrary (and whose effect is unknown), and using heuristics with known imperfections. All they can say is that certain articles are more likely to be plagiarism than others: saying that some may be plagiarism and some may not requires to introduce an arbitrary cut-off. In spite of all this, the conclusion made public is that ``the search turned up 677 examples of possible plagiarism'' \cite{Giles-06}.

\subsection{Trifle and punishment}
When there exist only two categories, labeling becomes crucial. Anything labeled plagiarism will be treated in a similar fashion. In the case of the plagiarized background section, \citet{Loui-02} reports that ``the funding agency insisted that the university conduct an investigation and issue a finding of plagiarism; the respondent was barred from receiving funding for one year.'' He disagreed with this sanction: ``the debarment from funding was excessive. The copying of a background section, without expropriation of others' original ideas, is a venial sin when compared with deliberate deception and theft of ideas.'' (Ginsparg, cited in Ref.~\onlinecite{Brumfiel-07}, agrees that ``although such practices are ethically questionable, it is inappropriate to be overly draconian.'')
It was important that the finding be of `plagiarism:' had it been labeled ``poor academic practice''~\cite{Burkill-04} for instance, the debarment from funding would have been perceived as disproportionate. But with the `plagiarism' label, this becomes acceptable. Bear in mind that arbitrarily arresting people and detaining them without any known charges is not outrageous when they are labeled terrorists~--- in both cases the justifications for the punishment is the wrongness of a label rather than the wrongness of anything actual. Nevertheless, ``referencing practices should not be based on the fear of being labeled a plagiarist''~\cite{Haramati-94}.

But acknowledging differences of wrongness is not the same as endorsing plagiarism: one can say that certain forms of copying from the works of others are venial without implying that genuine plagiarism is acceptable (likewise, the concept of self-defense is an not endorsement of murder). The blanket term `plagiarism' is as useless (and potentially dangerous) as `killing' when discussing punishment: it bundles deeds very different in nature and importance into a single label and treats them similarly because they have the same label rather than because they are similarly wrong.

\subsection{Students write to avoid plagiarism}
As \citet{Hunt-02} points out, ``scholars ---writers generally--- use citations for many things [\,\ldots]. They do not use them to defend themselves against potential allegations of plagiarism. [\,\ldots] Typically, the scholars are achieving something positive; the students are avoiding something negative.'' In fact this is what is expected of them ---students are taught ``writing to avoid plagiarism''~(a section heading in Ref.~\onlinecite{Burkill-04})--- even though avoiding plagiarism plainly cannot be a proper goal. When one walks, one does not walk to avoid falling: one walks to go somewhere. And one does not spend one's time thinking about not falling. Teaching students `writing to avoid plagiarism' is like teaching `walking to avoid falling': it misses the purpose of writing. 

When avoiding plagiarism is the goal, anything that can contribute to it is a valuable skill. Students are thus taught `proper' paraphrasing. \Citet{Hunt-02} comments an example of paraphrase
: ``What is clearest about this is that the writer of the second paragraph has no motive for rephrasing the passage other than to put it into different words.'' As \citet{McCutchen-94} notes, ``it is a strange crime that can be exorcised by the spell of replacing one set of words by another set that means the same thing.''
When the goal is the creation of new knowledge, on the other hand, paraphrasing is about worthless, as it is concerned with repeating old stuff.

The focus of \citet{Burkill-04} on ``regulations'' and on ``penalties'' for ``ignoring academic conventions'' indicates that to them the main reasons for students to avoid plagiarism are obedience and avoidance of penalties. These may be suitable driving forces in the training of dogs, but not for students (not to mention scholars). 
\Citet{Kohn-96} notes that even when students ``are `successfully' reinforced or consequenced into compliance, they will likely feel no commitment to what they are doing, no deep understanding of the act and its rationale, no sense of themselves as the kind of people who would want to act this way in the future.'' As Nietzsche~\cite[\S~97]{Nietzsche-dawn} pointed out, ``submission to morality can be slavish [\,\ldots]\ or resigned [\,\ldots]\ or an act of desperation, like submission to a prince: in itself it is nothing moral''~--- dread and honesty can both be driving forces for not plagiarizing. Turning the most trivial deed into plagiarism and focusing on penalties can foster only the former. 
\Citet{Burkill-04} also write that ``it may sound rather harsh, but the sooner you [the student] get to grips with all of this, the more confident you will feel as a member of the institution you have joined'': at no point is it claimed that such rules are justified, it is just that the faster the students obey the less they will get hurt. Students are not expected to accomplish anything positive, only to avoid punishment. 
Kohn rejected this Pavlovian view of education in the case of elementary school children; perhaps we could also reject it in high schools and universities.

\end{document}